# Artificial Intelligence Assistance Significantly Improves Gleason Grading of Prostate Biopsies by Pathologists

Wouter Bulten[1], Maschenka Balkenhol[1], Jean-Joël Awoumou Belinga[2], Américo Brilhante[3], Aslı Çakır[4], Xavier Farré[5], Katerina Geronatsiou[6], Vincent Molinié[7], Guilherme Pereira[8], Paromita Roy[9], Günter Saile[10], Paulo Salles[11], Ewout Schaafsma[1], Joëlle Tschui[12], Anne-Marie Vos[1], Hester van Boven[13], Robert Vink[14], Jeroen van der Laak[1,15], Christina Hulsbergen-van de Kaa[14], Geert Litjens[1]

1. Radboud University Medical Center, Radboud Institute for Health Sciences, Department of Pathology, Nijmegen, The Netherlands.
2. Faculty of Medicine and Biomedical Sciences, University of Yaounde 1, Department of Morphological Sciences and Anatomic Pathology, Cameroon
3. Salomão Zoppi Diagnostics/DASA, São Paulo, Brazil
4. Istanbul Medipol University, School of Medicine, Pathology Department, Istanbul, Turkey
5. Department of Health. Public Health Agency of Catalonia. Lleida, Catalonia, Spain
6. Centre de Pathologie, Hopital Diaconat Mulhouse, France
7. Pathology department, Aix en Provence Hospital, France
8. Histo Patologia Cirúrgica e Citologia, João Pessoa-PB, Brazil
9. Tata Medical Center, Department of Pathology, Kolkata, India
10. labor team w ag, Abteilung für Histopathologie und Zytologie, Goldach SG, Switzerland
11. Instituto Mário Penna, Belo Horizonte, Brazil
12. Medics Pathologie, Bern, Switzerland
13. Department of Pathology, Antoni van Leeuwenhoek Hospital, The Netherlands Cancer Institute, Amsterdam, The Netherlands
14. Laboratory of Pathology East Netherlands, Hengelo, The Netherlands
15. Center for Medical Image Science and Visualization, Linköping University, Linköping, Sweden

**Corresponding Author**

Wouter Bulten

wouter.bulten@radboudumc.nl

Radboud University Medical Center

Department of Pathology

Postbus 9101

6500 HB Nijmegen

The Netherlands

**Abstract**

While the Gleason score is the most important prognostic marker for prostate cancer patients, it suffers from significant observer variability. Artificial Intelligence (AI) systems, based on deep learning, have proven to achieve pathologist-level performance at Gleason grading. However, the performance of such systems can degrade in the presence of artifacts, foreign tissue, or other anomalies. Pathologists integrating their expertise with feedback from an AI system could result in a synergy that outperforms both the individual pathologist and the system. Despite the hype around AI assistance, existing literature on this topic within the pathology domain is limited. We investigated the value of AI assistance for grading prostate biopsies. A panel of fourteen observers graded 160 biopsies with and without AI assistance. Using AI, the agreement of the panel with an expert reference standard significantly increased (quadratically weighted Cohen's kappa, 0.799 vs 0.872; p=0.018). Our results show the added value of AI systems for Gleason grading, but more importantly, show the benefits of pathologist-AI synergy.

# Introduction

The biopsy Gleason score is the most important prognostic marker for prostate cancer patients.[1] However, it has been shown that Gleason grading suffers from significant inter- and intraobserver variability.[2,3] Specialized uropathologists show higher concordance rates,[4] but such expertise is not always available. Artificial intelligent (AI) systems based on deep learning have achieved pathologist-level performance in Gleason grading.[5-8] We investigated whether pathologists supported by a deep learning system improve in Gleason grading of prostate biopsies.

Pathologists assess the Gleason grade of a prostate biopsy through microscopic assessment of tissue stained with hematoxylin and eosin (H&E). Based on the morphological pattern of the tumor, a grade between one and five is assigned, with one being the lowest and five the highest. For biopsies, the Gleason score is the sum of the two most common patterns, e.g., 3+5=8. If a higher tertiary pattern is present, this is used instead of the secondary pattern. Patterns 1 and 2 are not reported anymore for biopsies.[9]

Recently, grade groups were introduced to improve the reporting of Gleason grading by assigning the Gleason score to one of five prognostic groups.[10] These groups are directly based on the Gleason score; 3+3 and lower go to group 1, 3+4 to group 2, 4+3 to group 3, 3+5, 5+3 and 4+4 to group 4, and higher scores to group 5. While the introduction of grade groups showed clinical value and increased interpretability of the tumor grade for patients, it has not improved the inter- and intraobserver variability.[5,11]

Deep learning has shown promise in many medical fields,[12] and the introduction of digital pathology allows for AI-based diagnostics in pathology.[13] For prostate cancer, methods based on deep learning have been developed for tumor detection,[14,15] grading of prostatectomies,[5] tissue microarrays[6] and biopsies.[7,8,16] In multiple studies, such deep learning systems showed pathologist-level performance, within the limits of the study setup.[5,7,8]

While deep learning systems have shown to achieve high performances on grading tasks, evidence of the merit of such systems when embedded in the pathologist's workflow is limited. Deep learning systems can be viewed as a new tool for pathologists to use in their diagnostic process and should also be evaluated as such. Additionally, regardless of the merits, most developed systems are also constrained by significant limitations that affect the performance and can lower the diagnostic power. Within histopathology, the presence of non-prostate tissue, atypical tissue patterns, ink on a slide, fixation, scanning and cutting artifacts, or the presence of rare cancer subtypes can dramatically affect a system's assessment of tissue. Many of these errors, especially those caused by artifacts, are easily spotted by a human observer.

Studies combining experts' opinions with feedback from automated systems have mainly been performed outside of the field of pathology; for example on the task of breast cancer detection in mammography.[17] For pathology, on the task of cancer metastasis detection in lymph nodes, the sensitivity of detection of micrometastases increased, and overall case reading time went down as a result of AI support.[18] On the task of mitosis counting, AI-generated hotspots improved reproducibility between readers.[19] However, most of these studies focus either on computer-aided detection (CADe) or diagnosis (CADx). For prognostic measures, such as Gleason grading of prostate biopsies, there is, to the best of our knowledge, no such study as of yet.

In a previous study, we developed a fully automated deep learning system for grading prostate cancer.[7] The deep learning system was trained on a large dataset of prostate biopsies and achieved pathologist-level performance, both in determining the grade group and in stratifying patients in relevant risk categories. As part of the initial validation of the system, it's performance was compared to a panel of pathologists in an observer experiment. The deep learning system outperformed 10 out of 15 observers on determining the grade group.

In this study, we investigate the value of AI-assisted reading by pathologists for Gleason grading of prostate biopsies. We compare the diagnostic performance of pathologists with and without the assistance of the deep learning system. We show that the integration of AI feedback in the diagnostic process has merits and that a synergy between the pathologist and AI system achieves the best performance overall.

## Results

### The observer panel

We invited 15 pathologists (board certified or residents) who participated in an earlier study on automated Gleason grading[7] to perform this observer experiment. Two panel members dropped out due to other obligations or a lack of time; one additional pathology resident joined the current study. In total, the observer panel consisted of 14 members (11 certified pathologists and three pathology residents), originating from 12 independent labs and eight countries. All panel members had prior experience with Gleason grading, though with varying amounts of experience.

### Dataset under review and reference standard

From the test set that was previously used to evaluate our deep learning system,[7] a set of 160 cases was selected to be reviewed by the panel. All cases under review had been graded by three uropathologists with extensive (>20 years) experience in Gleason grading, and their consensus opinion set the reference standard. Grading was performed according to protocol and as part of our previous study.[7] Of the selected cases, 100 cases were already graded by the panel as part of the previous study and were reused for the current study; the remaining 60 cases were unseen to act as controls, to measure the potential effect of a second-read on the original cases. The complete set of 160 cases for the AI-assisted read in the present study consisted of 30 (19%) benign cases, 22 (14%) cases with grade group 1, 26 (16%) cases with grade group 2, 32 (20%) cases with grade group 3, 20 (13%) cases with grade group 4 and 30 (19%) cases with grade group 5.

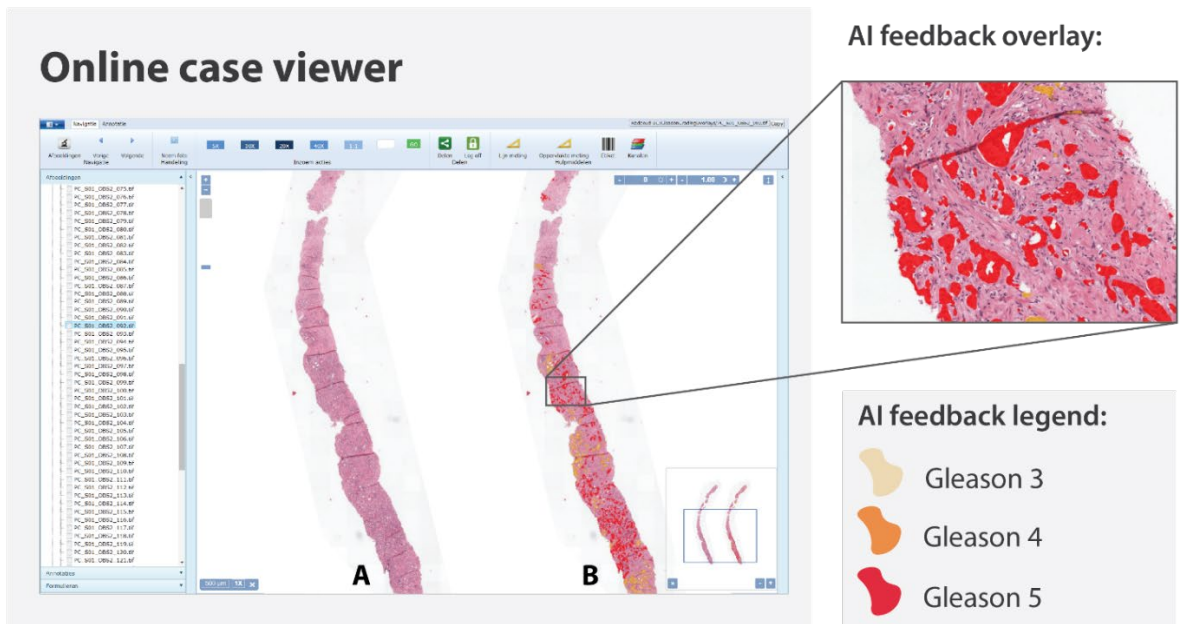

**Figure 1.** Overview of the viewer used in the observer experiment. Both the original biopsy (A) and the biopsy with the AI overlay (B) are presented to the pathologist. Each individual tumor gland is marked by the deep learning system in the overlay. The case-level grade group was supplied to the panel as part of their (separate) grading form.

AI-assisted Gleason grading

Grading by the panel members was performed using an online viewer (Figure 1), after which the panel members filled in a questionnaire on the grading process. The viewer showed the original biopsy in addition to a copy with a color overlay where different AI-predicted growth patterns were highlighted using different colors. In addition, the readers were presented with the automatically predicted biospy-level grade group for every biopsy. Five out of 14 (36%) panel members predicted that they scored somewhat better in comparison to the first read. Of these five, three out of 14 (21%) expected a performance increase due to being more experienced in viewing cases using the online viewer, and two out of 14 (14%) because of the AI assistance. The majority of the panel members (8 out of 14, 57%) indicated they did not expect a performance increase as a result of the AI assistance, while one pathologist (1 out of 14, 7%) expected to have scored somewhat lower.

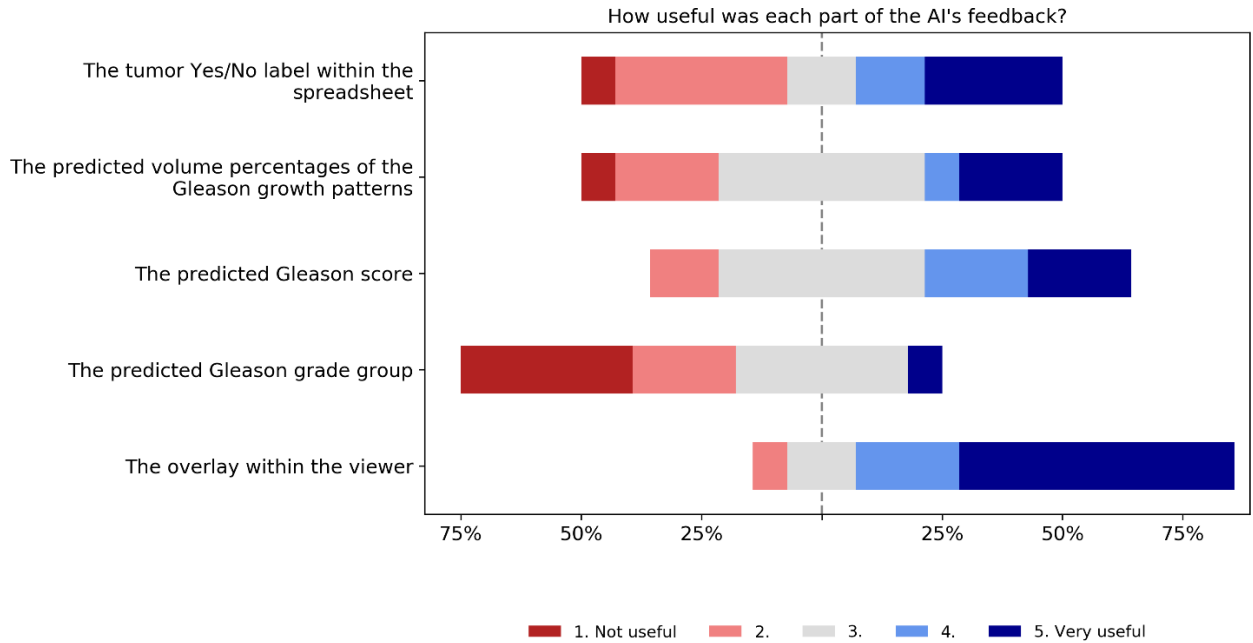

**Figure 2.** Survey results on the AI feedback. Panel members were asked to indicate how useful each part of the AI's feedback was on a five-point scale from "Not useful" to "Very useful."

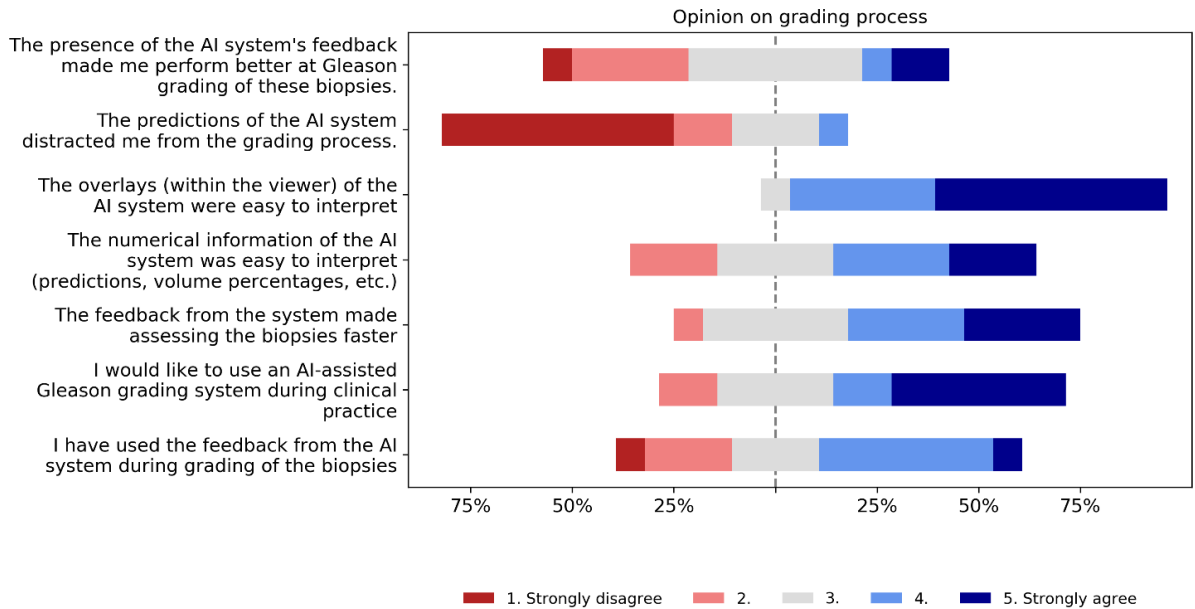

**Figure 3.** Survey results on the grading process. Panel members were asked to reflect on the grading process and answer questions on a five-point scale from "Strongly disagree" to "Strongly agree."

Eleven out of 14 (79%) panel members indicated that they used feedback from the AI system during grading. Of all the components of the AI feedback, the growth pattern overlay was determined to be the most useful and easy to interpret (Figure 2). The panel members indicated that the final grade group, as assigned by the system, was the least helpful. The majority of panel members noted that the AI assistance did not distract them from the grading process, but instead made grading the biopsies faster (Figure 3).

Performance of the panel with and without AI feedback

In the first read without AI assistance, the agreement with the reference standard (measured by the median quadratically weighted Cohen's kappa) for the panel was 0.799. In the second, AI-assisted read, the median kappa of the panel increased to 0.872 (9.14% increase), showing a significant increase in performance (Wilcoxon signed-rank test p=0.018, Figure 4). On the same dataset, the AI system in itself achieved a kappa score of 0.854. Excluding panel members who estimated that they improved due to viewing more cases (n=3) or excluding pathologists who indicated that they did not use the AI feedback (n=3), we found a comparable increase in median kappa from respectively 0.754 to 0.874 (p=0.041) and 0.754 to 0.870 (p=0.016).

Nine of the 14 (64%) panel members scored higher in the assisted read, while five (36%) panel members scored slightly lower, though with a maximum decrease in kappa score of 0.013. Of the five that scored lower, four already outperformed the AI system in the first read. The interquartile range of the panel's kappa values dropped from 0.113 to 0.073 in the second read (Figure 4).

In the first read, the kappa value of the AI system exceeded that of 10 out of the 14 (71%) panel members. In the AI-assisted read, only five of the panel members (36%) scored a kappa value below that of the AI system. The largest improvement was seen for panel members who had less than 15 years of experience (Figure 5). Of the panel members who scored lower than the AI system in the unassisted read (10 out of 14, 71%), nine scored higher in the assisted read (9 out of 10, 90%). None of the panel members who outperformed the AI in the unassisted read improved in the assisted read. On a group-level, the median performance of AI-assisted reads was higher than both that of the AI system itself and the unassisted reads.

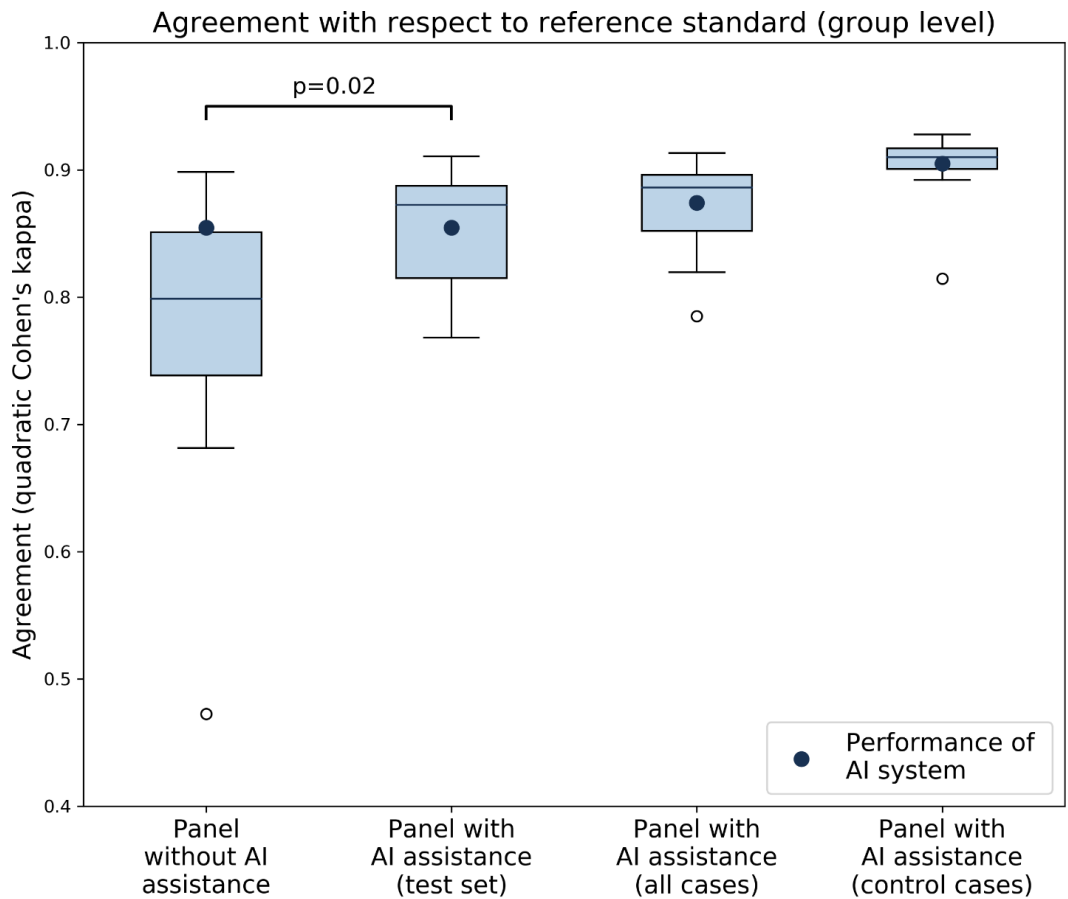

**Figure 4.** Panel performance with and without AI assistance. With AI assistance, the median performance of the group increased while the variability between panel members went down.

The agreement of the panel with the reference standard on the control cases was high, with a median kappa value of 0.910, and slightly higher in comparison to the test cases (kappa 0.872). The control cases were only viewed in the assisted read. The system's performance on the control cases was also higher, with a kappa value of 0.905 compared to 0.854 on the test cases.

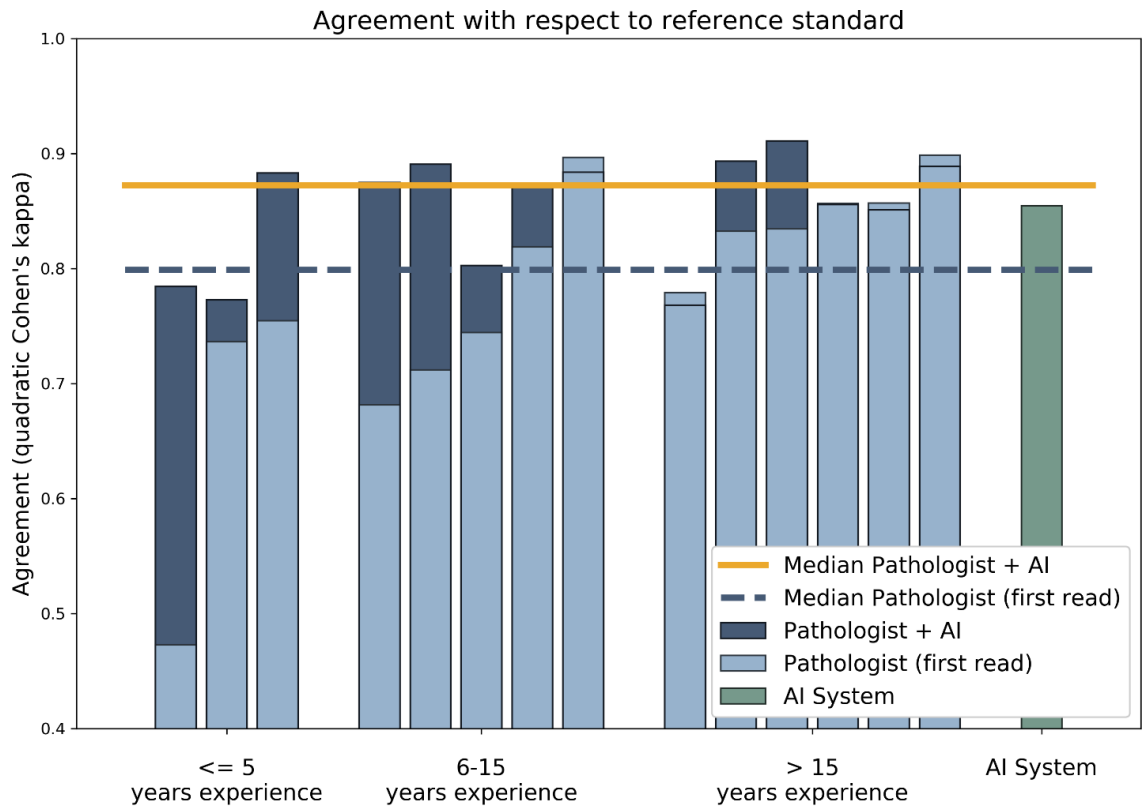

**Figure 5.** Individual performance of panel members shown for both the unassisted read (light blue) and assisted read (dark blue). A lower performance in the unassisted read is indicated with a line in the light blue bars. Pathologists are sorted based on experience level and the kappa value of the unassisted read. The performance of the standalone AI system is shown in green. In the unassisted read, the AI system outperforms the group. In the assisted read, the median performance of the group is higher than of the standalone AI system.

## Discussion

To the best of our knowledge, this study was the first to explore the possible merits of AI assistance on histological tumor grading. We showed that AI assistance improves pathologists' performance at Gleason grading of prostate biopsies. Measured through the agreement with an expert reference standard, the read with AI assistance resulted in a significant increase in performance (quadratically weighted Cohen's kappa, 0.799 vs 0.872). Moreover, the variance between panel members' performance decreased, resulting in overall more consistent grading. Reduced observer variability of Gleason grading is highly desirable, as it yields a stronger prognostic marker for individual patients and reduces the effect of the diagnosing pathologist on potential treatment decisions.

In the unassisted read, the AI system outperformed 10 out of 14 pathologists, and this dropped to only 5 out of 14 in the second-read with AI assistance. Pathologists assisted by the AI system not only improved compared to unassisted reads but also achieved higher median performance than the standalone AI. These results show that there is an added benefit of pathologists using AI assistance as a supportive tool during diagnosis. Especially in geographic regions where the number of pathologists is limited or subspecialized pathologists are not available, AI systems such as ours can support pathologists in achieving higher grading accuracy and consistency.

The most substantial increase in performance was seen for panel members who initially scored lower than the AI system. Most of the pathologists with more than 15 years of experience, who often outperformed the AI system in the unassisted read, scored comparably in both reads. While no performance gain was found for these pathologists in terms of diagnostic accuracy, the pathologists indicated that the use of the AI system led to faster grading. However, in this study, we did not directly measure the time taken per case, so this remains an interesting direction for future research.

Through a questionnaire, we investigated the pathologists' experiences when using the system. One of the design goals of the deep learning system was to support the workflow of pathologists. The system was developed to give feedback on multiple abstraction levels, with the grade group giving an overall assessment and the overlay more detailed feedback. We assumed that the precise gland-level segmentations of the tumor and growth patterns could support pathologists in quickly assessing glandular regions and assisting in volume measurements. Almost all pathologists indicated that the AI system's overlay was useful, and, based on the questionnaire, was the most used feature of the system. AI assistance through these overlays can be seen as another tool for pathologists that gives feedback on a glandular level, comparable to, e.g., immunohistochemistry, and gives direct support for the systems' case-level prediction. The overlays allow pathologists to combine their expertise with the added feedback of the system to determine the final grade.

The AI feedback was also given on a case-level through volume measurements, the Gleason score, and the grade group. Of all features, panel members rated the biopsy level grade group as the least useful. Given that the grade group is directly computed from the Gleason score, it can be seen as redundant information in the feedback.

While the results of this observer experiment are promising, some limitations have to be addressed. First, we cannot entirely exclude that factors outside of the AI feedback influenced the pathologists' performance, both positively and negatively. While pathologists did not receive any feedback between the two reads, more experience with viewing cases digitally, the viewer, or in Gleason grading itself could have some influence on the results. Though, we believe that the influence of a second-read is small for several reasons: the majority of pathologists predicted that they scored the same, a significant increase was still found when excluding pathologists who indicated more experience, and the performance on the unseen control cases was also high.

Secondly, pathologists were not extensively trained or instructed to use the AI system and were free to use the system in any way during grading. All cases that were graded were also included in the analysis, and we did not allow for a training phase. Pathologists can benefit from an understanding of the global properties of an AI system when such a system is introduced in their grading process; this includes the system's limitations, its tendency to over- or under grade, and the overall design goal.[20] A training phase at the start of the observer experiment could have increased the use, understanding, and effectiveness of the AI feedback during the grading process and might have led to further increased performance by using the AI system.

Last, we performed our observer study using a limited set of 100 test and 60 control cases from a single center. The difference in performance between the test set (median kappa 0.872) and the control set (median kappa 0.910) shows that case difficulty can influence overall results. To further strengthen our conclusions, the AI feedback could be tested with a larger set of cases from different centers.

Our results open up abilities for future research. First, most pathologists indicated that the AI feedback made assessing the biopsies faster, but this was not measured quantitatively in the current study. For clinical applications, where reducing the workload is an important topic, saving time through AI assistance is of great interest. Second, during routine diagnostics, pathologists will assess multiple biopsies from a single patient and integrate all case information into the final diagnosis. In the current study, we limited the grading process to single biopsies per case, both for the system as for the pathologist. An interesting avenue for future research would be to investigate AI assistance on a patient-level, which allows for new approaches such as automatically prioritizing slides.

## Methods

### Collection of the dataset and setting the reference standard

In a previous study,[7] we developed a deep learning system to grade prostate biopsies using the Gleason grading system. To train this system we collected a dataset of 5759 H&E-stained biopsies corresponding to 1243 patients. All biopsy procedures were performed as part of routine diagnostics at the Radboud University Medical Center between 2012 and 2017. For the study, the H&E stained glass slides of the biopsies were digitized at 20x magnification (pixel resolution 0.24µm) using a 3DHistech Pannoramic Flash II 250 scanner and subsequently anonymized. The need for informed consent was waived by the local ethics review board (IRB number 2016-2275).

Of the dataset, 550 biopsies were excluded from model development and used as an independent test set to evaluate the deep learning system. Patients that were included in this test set were independent of the patients in the training set. Given the inter-observer variability of Gleason grading, validation of the system required a consensus reference standard. In the first round, three expert pathologists (C.H.v.d.K., R.V., H.v.B.) with a subspecialty in uropathology individually graded the cases in the test set using the ISUP 2014 guidelines.[21] For some cases the majority vote was taken: cases with an agreement on grade group but a difference in Gleason pattern order, e.g., 5+4 versus 4+5; cases with an equal grade group but a disagreement on Gleason score; and cases for which two pathologists agreed while the third had a maximum deviation of one grade group. Cases with a disagreement on malignancy were always flagged. In the second round, cases that had no agreement were presented to the pathologist who deviated the most from the other two. Additional to the pathologist's score, the scores of the two other pathologists were shown anonymously. Finally, biopsies without agreement after two rounds were discussed in a consensus meeting.

### Observer panel and case selection

Part of the first study was a comparison of the deep learning system to a panel of pathologists. Of the full test set, 100 cases were selected and presented to a panel of 13 external pathologists and two pathologists in training. Of these 100 cases, 20 benign cases were selected by one of the expert pathologists (C.H.v.d.K.) to cover diverse tissue patterns, including those which are considered challenging prostate cancer mimics such as granulomatous inflammation. The other 80 cases were sampled uniformly based on the grade group determined by the same pathologist. The panel was asked to grade all biopsies through an online viewer PMA.view (Pathomation, Berchem, Belgium) following the ISUP 2014 guidelines. No time limit was set for the grading process.

All pathologists that participated in the first study were invited to participate in the present study. One additional pathologist in training, who showed interest in the first study but was not able to grade all biopsies before submission of the previous paper, was also asked to join the current study.

We included all 100 biopsies from the first study, as these were already graded by the pathologists in the panel. Additionally, we extended the dataset with 60 new cases from the original test set, all of which were unseen by the panel members. These new unseen cases were used as control cases to measure the potential effect of a second-read on the original cases. One of the expert pathologists who set the reference standard (C.H.v.d.K.), selected ten benign cases manually, again controlling for a broad range of tissue patterns. The remaining fifty cases were sampled uniformly over the five grade groups, based on the consensus label. All 160 biopsies were shuffled and assigned new identifiers.

### Feedback of the AI system

We processed each biopsy in the dataset using the deep learning system,[7] resulting in a prediction of the volume percentages of each growth pattern (if present), the Gleason score, and the grade group per biopsy. Besides a numerical prediction, the system also generated an overlay that outlined malignant glands: Gleason 3 in yellow, Gleason 4 in orange and Gleason 5 in red. For the current experiment, we chose not to highlight detected benign tissue. The overlays were preprocessed by a connected components algorithm to remove small artifacts and to ensure that each gland was assigned to a single growth pattern. Preprocessing was done fully automatically without manual review.

### Second-read with AI assistance

The 160 biopsies were made available to the panel of pathologists through the same online viewer as during the first read. The time between the first and second read was at least three months. Each biopsy could be viewed at a maximum pixel spacing of 0.24µm (roughly equivalent to 40x objective magnification). Next to the original biopsy, we showed an exact copy of the biopsy with the algorithm's prediction as an overlay (Figure 1). This overlay could be used to assess the tissue that the algorithm had flagged as malignant. To complement the overlay, we also supplied the numerical output of the deep learning system to the panel, including the predicted volume percentages, the presence of tumor (yes/no), the Gleason score, and the grade group.

We asked each panel member to report: whether a biopsy contained tumor, presence of Gleason growth patterns, volume percentages of present patterns, and the grade group. After grading a case, the panel members had to indicate whether they thought the system's prediction influenced their assessment.

No time restriction was given per case, but we asked each pathologist to complete all 160 cases within eight weeks. Each panel member was instructed to review the cases individually without consulting colleagues. Panel members had no access to the cases from the previous experiment nor to the grades that they assigned. Between the first and second read, no performance indication or feedback was given to panel members with respect to the reference standard. When all cases were graded, each panel member was asked to fill in a questionnaire regarding the process and feedback from the deep learning system.

## Statistical analysis

After all panel members completed the grading of the biopsies, we compared their scores to the consensus reference standard. Cohen's kappa with quadratic weights was used as the primary metric of performance. On a group-level, we used the median kappa as the metric to account for outliers.

To compare reading cases with or without AI assistance we conducted a statistical analysis, using the difference in kappa between the two reads as the test statistic. A Shapiro-Wilks test for normality was performed to show that the data was not normally distributed. To compare the difference in kappa scores, we performed a Wilcoxon signed-rank test on the paired kappa values. The test statistic was computed using the grades of the 100 cases that were used in both reads.

## Acknowledgments

This study was financed by a grant from the Dutch Cancer Society (KWF), grant number KUN 2015-7970. The authors would like to thank Steven Teerenstra for contributing to the statistical analysis.

## Author contributions

W.B. performed data selection, experiments, analyzed the results, and wrote the manuscript. M.B., J.A.B., A.B., A.C., X.F., K.G., V.M., G.P., P.R., G.S., P.S., E.S., J.T. and A.V. participated in the panel and graded all cases. H.v.B, R.V. and C.H.-v.d.K. graded all cases and set the reference standard. G.L. and J.v.d.L. supervised the work and were involved in setting up the experimental design. All authors reviewed the manuscript and agreed with its contents.

## Competing interests

W.B. reports grants from the Dutch Cancer Society, during the conduct of the study. J.v.d.L. reports personal fees from Philips, grants from Philips, personal fees from ContextVision, personal fees from AbbVie, grants from Sectra, outside the submitted work. G.L. reports grants from the Dutch Cancer Society, during the conduct of the study; grants from Philips Digital Pathology Solutions, personal fees from Novartis, outside the submitted work. M.B., J.A.B., A.B., A.C., X.F., K.G., G.P., P.R., G.S., P.S., J.T., H.v.B., R.V. and C.H.-v.d.K. have nothing to disclose.

## References

1  Epstein, J. I. An Update of the Gleason Grading System. *J. Urol.* **183**, 433-440, doi:10.1016/j.juro.2009.10.046 (2010).
2  Allsbrook, W. C. *et al.* Interobserver reproducibility of Gleason grading of prostatic carcinoma: General pathologists. *Hum. Pathol.* **32**, 81-88, doi:10.1053/hupa.2001.21135 (2001).
3  Egevad, L. *et al.* Standardization of Gleason grading among 337 European pathologists. *Histopathology* **62**, 247-256, doi:10.1111/his.12008 (2013).
4  Allsbrook, W. C. *et al.* Interobserver reproducibility of Gleason grading of prostatic carcinoma: Urologic pathologists. *Hum. Pathol.* **32**, 74-80, doi:10.1053/hupa.2001.21134 (2001).
5  Nagpal, K. *et al.* Development and Validation of a Deep Learning Algorithm for Improving Gleason Scoring of Prostate Cancer. *npj Digital Medicine*, doi:10.1038/s41746-019-0112-2 (2018).
6  Arvaniti, E. *et al.* Automated Gleason grading of prostate cancer tissue microarrays via deep learning. *Sci. Rep.* **8**, 1-11, doi:10.1038/s41598-018-30535-1 (2018).
7  Bulten, W. *et al.* Automated deep-learning system for Gleason grading of prostate cancer using biopsies: a diagnostic study. *The Lancet Oncology* **21**, 233-241, doi:https://doi.org/10.1016/S1470-2045(19)30739-9 (2020).
8  Ström, P. *et al.* Artificial intelligence for diagnosis and grading of prostate cancer in biopsies: a population-based, diagnostic study. *The Lancet Oncology* **21**, 222-232, doi:https://doi.org/10.1016/S1470-2045(19)30738-7 (2020).

9       Epstein, J. I. *et al.* The 2005 International Society of Urological Pathology (ISUP) Consensus Conference on Gleason Grading of Prostatic Carcinoma. *Eur. Urol.* **49**, 758-759, doi:10.1016/j.eururo.2006.02.007 (2006).
10      Epstein, J. I. *et al.* A Contemporary Prostate Cancer Grading System: A Validated Alternative to the Gleason Score. *Eur. Urol.* **69**, 428-435, doi:10.1016/j.eururo.2015.06.046 (2016).
11      Ozkan, T. A. *et al.* Interobserver variability in Gleason histological grading of prostate cancer. *Scandinavian Journal of Urology* **50**, 420-424, doi:10.1080/21681805.2016.1206619 (2016).
12      Litjens, G. *et al.* A survey on deep learning in medical image analysis. *Med. Image Anal.* **42**, 60-88, doi:10.1016/j.media.2017.07.005 (2017).
13      Niazi, M. K. K., Parwani, A. V. & Gurcan, M. N. Digital pathology and artificial intelligence. *The Lancet Oncology* **20**, e253-e261, doi:https://doi.org/10.1016/S1470-2045(19)30154-8 (2019).
14      Litjens, G. *et al.* Deep learning as a tool for increased accuracy and efficiency of histopathological diagnosis. *Sci. Rep.* **6**, 26286, doi:10.1038/srep26286 (2016).
15      Campanella, G. *et al.* Clinical-grade computational pathology using weakly supervised deep learning on whole slide images. *Nat. Med.* **25**, 1301-1309, doi:10.1038/s41591-019-0508-1 (2019).
16      Lucas, M. *et al.* Deep learning for automatic Gleason pattern classification for grade group determination of prostate biopsies. *Virchows Arch.*, doi:10.1007/s00428-019-02577-x (2019).
17      Rodríguez-Ruiz, A. *et al.* Detection of Breast Cancer with Mammography: Effect of an Artificial Intelligence Support System. *Radiology* **290**, 305-314, doi:10.1148/radiol.2018181371 (2018).
18      Steiner, D. F. *et al.* Impact of Deep Learning Assistance on the Histopathologic Review of Lymph Nodes for Metastatic Breast Cancer. *The American journal of surgical pathology* **42**, 1636-1646, doi:10.1097/PAS.0000000000001151 (2018).
19      Balkenhol, M. C. A. *et al.* Deep learning assisted mitotic counting for breast cancer. *Lab. Invest.* **99**, 1596-1606, doi:10.1038/s41374-019-0275-0 (2019).
20      Cai, C. J., Winter, S., Steiner, D., Wilcox, L. & Terry, M. Hello AI: Uncovering the Onboarding Needs of Medical Practitioners for Human-AI Collaborative Decision-Making. *Proceedings of the ACM on Human-Computer Interaction* **3**, 104 (2019).
21      Epstein, J. I. *et al.* The 2014 International Society of Urological Pathology (ISUP) Consensus Conference on Gleason Grading of Prostatic Carcinoma. *The American Journal of Surgical Pathology* **40**, 1, doi:10.1097/pas.0000000000000530 (2015).